\let\oldyear\year
\documentclass[doublecol]{epl2} 

\let\year\oldyear
\usepackage{amsmath}
\usepackage{amssymb}
\usepackage{graphicx}
\usepackage{dcolumn}
\usepackage{bm}
\usepackage{pgfplots}
\usepackage{wasysym}
\usepackage[colorinlistoftodos]{todonotes}
\pgfplotsset{width=10cm,compat=1.9}
\usepackage{physics}
\usepackage{caption}
\usepackage{yfonts}
\usepackage{euscript}
\usepackage{subcaption}
\usepackage{multirow}
\usepackage{booktabs}

\newtheorem{definition}{Definition}

\title{Enhanced Forman curvature and its relation to Ollivier curvature}

\author{Philip Tee\inst{1,2} \and C.A. Trugenberger\inst{3}}
\shortauthor{P.Tee \etal}

\institute{                    
  \inst{1} The Beyond Center for Fundamental Science, Arizona State University, Tempe AZ\\
  \inst{2} Department of Informatics, University of Sussex, Falmer, UK.\\
  \inst{3} SwissScientific Technologies SA, rue du Rhone 59, CH-1204 Geneva, Switzerland
}
\pacs{02.40.Sf} {Manifolds and cell complexes}
\pacs{04.60.-m}{Quantum gravity}
\pacs{02.10.Ox} {Combinatorics; graph theory}

\abstract{
Recent advances in emergent geometry and discretized approaches to quantum gravity have relied upon the notion of a discrete measure of graph curvature.
We focus on the two main measures that have been studied, the so-called Ollivier-Ricci and Forman-Ricci curvatures.
These two approaches have a very different origin, and both have advantages and disadvantages.
In this work we study the relationship between the two measures for a class of graphs that are important in quantum gravity applications.
We discover that under a specific set of circumstances they are equivalent, opening up the possibility of replacing the more fundamental Ollivier-Ricci curvature by the computationally more accessible Forman-Ricci curvature in certain applications to 
models of emergent spacetime and quantum gravity.
}

\begin{document}

\maketitle

\section{Introduction}
Curvature is a fundamental concept in general relativity. 
In the absence of matter, the Einstein equations determine the stationary points of the Einstein-Hilbert action, the integral of the scalar curvature over the manifold. 
The Einstein-Hilbert action is, however, perturbatively non-renormalizable. 
Inspired by the successes of lattice gauge theory, one of the avenues that has been explored to cure this problem is to get rid of the associated ``infinities" by regularizing spacetime in terms of simplicial complexes and search for an ultraviolet (UV) fixed-point that defines quantum gravity non-perturbatively. 
In general, this dynamical triangulation program failed \cite{ambjorn1997quantum, ambjorn2013euclidian}, although there are indications that fine-tuning a parameter in a version of the model with a non-uniform measure may provide a way out \cite{laiho2017lattice}. The program does much better, however when a preferred foliation is assumed, so that one is essentially discretizing a Lorentz manifold. 
The resulting {\it causal} dynamical triangulations program has a rich structure with a much better scaling behaviour \cite{ambjorn1997quantum, loll2019}. 
An alternative approach is causal set theory \cite{bombelli1987space}, in which spacetime is considered as fundamentally discrete, with the structure of a locally finite poset describing the causal structure of spacetime (for a recent review see \cite{surya2019causal}). 

Simplicial complexes are still piece-wise flat chunks of spacetime. Recently an approach to discretize geometry by much ``wilder" structures, like random graphs (for a review see \cite{albert2002statistical}) has been proposed \cite{trugenberger2017combinatorial}. The idea of this combinatorial quantum gravity approach (CQG) is for geometric manifolds to emerge from random graphs in a continuous network transition, for which there is indeed strong evidence, e.g. the divergence of the correlation length and of the specific heat at the critical point \cite{kelly2019self,kelly2021emergence,tee2020dynamics}. Contrary to previous discrete approaches, CQG is agnostic as to the signature of the metric on the emerged manifolds. Depending on the specific instance these can support a Riemann metric or both a Riemann and a Lorentz metric. 

The commonality among these approaches is that a discrete notion of curvature is needed. For (causal) dynamical triangulation this is the original Regge curvature \cite{regge1961general,cheeger1984curvature} based on angle deficits. Angle deficits are also the basis for a generalized Regge curvature needed when random triangulations are admitted \cite{carfora2002modular}. Random triangulations are still geometric objects, even if fluctuating, and thus very different from the purely combinatorial objects in CQG. For causal sets, the appropriate construction is the Benincasa-Dowker curvature \cite{benincasa2010scalar}. 
For random graphs, several notions of purely combinatorial Ricci curvature have been advanced. 
In the original proposal \cite{trugenberger2017combinatorial} the Ollivier combinatorial Ricci curvature \cite{ollivier2007ricci, ollivier2009ricci, ollivier2011visual} was used. This has been recently shown \cite{van2021ollivier, kelly2021convergence} to converge to standard continuum Ricci curvature on random geometric graphs \cite{penrose2003random} and is thus a genuine candidate for a discretization of general relativity. A simplified variant of Ollivier curvature has also been recently introduced in \cite{klitgaard2018introducing,klitgaard2018implementing,klitgaard2020round}. 
Combinatorial curvature is distinctly different to Regge calculus \cite{regge1961general} and other forms of curvature more commonly used in EDT or CDT, as the combinatorial measures of curvature are not defined geometrically.
Instead, both measures have a purely combinatorial origin, and have a more natural relationship with CQG.

Ollivier curvature does however, have a very intuitive interpretation from the geometric point of view, since it mimics on a graph the fundamental geometric definition of continuum Ricci curvature (see below). However not only one, but several other notions of graph curvature have been proposed (for a taxonomy with brief description of the rationale for each we refer to \cite{van2021ollivier}). Here we focus on another widely studied and used concept of discrete Ricci curvature, Forman curvature \cite{forman2002combinatorial, forman2003bochner,forman2004topics}. 

Forman curvature was defined originally for cell complexes, specifically CW complexes. It reduces to a graph curvature if this is considered as a 1-complex, so that only nodes ($0$-cells) and links ($1$-cells) are considered. It can be augmented, by considering the graph as a $2$-complex, so that closed cycles are considered as bounding a $2$-cell. However, the focus in prior work was on triangles \cite{samal2018comparative}. Discrete locality, however, requires that the action at one vertex depends on nearest and next-to-nearest neighbours, the discrete equivalent of first and second derivatives. Consequently even this augmented Forman curvature is not appropriate for applications to CQG, since  locality requires considering cycles of length up to $5$, as automatically realized in the Ollivier curvature. Here we generalize the Forman construction for graphs considered as $2$-complexes and we show that the resulting ``enhanced" Forman Ricci curvature matches the mean field Ollivier curvature on random graphs satisfying the independent short-cycle condition \cite{kelly2019self}. These are the main results of our paper. 
They suggest that the dependence of local combinatorial curvature measures on the numbers of short (up to length 5) cycles based on an edge is unique.

\section{Ollivier-Ricci Curvature}
\label{sec:or_curvature}

The Ollivier-Ricci (OR) curvature is a measure of graph curvature first introduced by Yann Ollivier \cite{ollivier2007ricci,ollivier2009ricci,ollivier2011visual}.
It tries to mimic in a discrete setting the geometric property of Ricci curvature as a description of how infinitesimal balls expand or shrink upon parallel transport. 
Two points in Riemannian space can be used to define a geodesic through them. Consider now balls of infinitesimal radius around these two points. In a positively curved space the average geodesic distance between all other points on the two balls is shorter than the distance along the geodesic between the centers. For negatively curved space the result is the opposite, and for flat space the distances are the same.

To carry over into discrete graphs the key observation is that a graph is a metric space, and the role of the balls can be undertaken by unit normalized probability distributions that have their support in the neighborhood of a vertex.
Geodesic distance is replaced by the so-called ``earth mover" or Wasserstein distance between balls centered around two vertices $i$ and $j$, denoted $b_i, b_j$.
In these balls, `` earth mass"" is assumed to be distributed according to a unit normalized measure $\mu_i$, and a transference plan measures the exchange of mass necessary to move the distributions from $b_i$ to $b_j$. The Wasserstein distance $W\left( \mu_i, \mu_j \right)$ is the optimal such transport plan. It is thus defined as
\begin{equation}
W\left( \mu_1, \mu_2 \right) = {\rm inf} \sum_{i\in b_i,j\in b_j} \xi(i,j)d(i,j) \ ,
\label{wasser}
\end{equation}
where $d(i,j)$ is the graph distance and the infimum has to be taken over all couplings (or transference plans) $\xi(i,j)$.
That is to say over all plans on how to transport a unit mass distributed according to $\mu_1$ around $i$ to the same mass distributed according to $\mu_2$ around $j$, 
\begin{equation}
\sum_j \xi (i,j) = \mu_1(i) \ , \qquad 
\sum_i \xi (i,j) = \mu_2(j) \ .
\label{transplan}
\end{equation}
The OR curvature of an edge $ij$ is then defined as 
\begin{equation}
\kappa^{\rm OR} (i,j)= 1- \frac{W\left( \mu_i, \mu_j \right)}{ d(i,j)} \ .
\label{olli}
\end{equation}
In the simplest realization the balls $b_i$ are chosen as unit balls, and the probability distributions $\mu_i$ are uniform on these unit balls. Note, however,that the convergence to continuum Ricci curvature on generic random geometric graphs requires larger, ``mesoscopic" balls, in order to ``feel" the curvature of the background manifold \cite{van2021ollivier, kelly2021convergence}. Unit balls, instead are sufficient in the flat case, when the curvature of the background manifold vanishes at all scales. 

\section{The independent short cycle condition}
\label{sec:shortcycles}

The Ollivier curvature is very intuitive but also very cumbersome to compute in general since one has to solve a linear programming problem for each edge. Remarkably, there exists a class of random graphs for which there exists a closed-form expression. Moreover, it turns out that these are exactly the most important graphs for applications to CQG, as we discuss in detail below.  

First of all we note that the simplest version of OR curvature on unit balls, due to its very definition depends only on the nearest and next-to-nearest neighbours of a vertex on the considered edge. 
This is a discrete version of locality; ``feeling" the influence of at least up to the next-to-nearest neighbours of a vertex is the minimum requirement of a combinatorial curvature notion that can be used to define discrete quantum gravity. 
It turns out that although the fundamental degrees of freedom in CQG are the edges, upon which curvature is defined, the physical degrees of freedom are actually cycles, or loops.
These cycles and loops represent a sort of discrete ``gauge principle" \cite{trugenberger2017combinatorial, kelly2019self,kelly2021emergence}. 
Locality implies that only triangles, squares and pentagons matter. 

\begin{figure}[h]
    \centering
    \begin{tikzpicture}[node distance=0.5cm]
	    \node [black] at (0,2) {\textbullet}; 
	    \node [black] at (0,0) {\textbullet}; 
	
	    \node [black] at (1,1) {\textbullet}; 
	    \node [black] at (2,2) {\textbullet}; 
	    \node [black] at (2,0) {\textbullet}; 
	    \node [black] at (2,2.75) {\textbullet}; 
	    \node [black] at (2,-0.75) {\textbullet}; 
	    \node [black] at (3.75,1) {\textbullet}; 

	    \draw [dashed] (0,0) -- (-0.4,-0.4);
	    \draw [dashed] (0,0) -- (0,-0.4);
	    \draw [dashed] (0,0) -- (0.4,-0.4);
	    \draw [dashed] (0,2) -- (-0.4,2.4);
	    \draw [dashed] (0,2) -- (0,2.4);
	    \draw [dashed] (0,2) -- (0.4,2.4);
	    \draw [-] (0,0) -- (0,2);
	    \draw [dashed,blue] (0,0) -- (2,0);
	    \draw [dashed,blue] (0,2) -- (2,2);
	    \draw [dashed,blue] (2,0) -- (2,2);
	    \draw [dashed,red] (0,0) -- (1,1);
	    \draw [dashed,red] (0,2) -- (1,1);
	    \draw [-] (0,0) -- (2,-0.75);
	    \draw [-] (0,2) -- (2,2.75);
	    \draw [-] (2,-0.75) -- (3.75,1);
	    \draw [-] (2,2.75) -- (3.75,1);
	
	    \node at (-0.3,0) {$i$};
	    \node at (-0.3,2) {$j$};
	    \node at (1.2,1) {$k$};
	    \node at (2.2,2) {$l$};
	    \node at (2.25,0) {$m$};
	    \node at (4,1) {$o$};
	    \node at (2,3) {$p$};
	    \node at (2,-1) {$n$};
	    \node at (0,2.75) {$N(j)$};
	    \node at (0,-0.75) {$N(i)$};
\end{tikzpicture}
\caption{We depict two vertices $i$ and $j$ and the connecting edge $e_{ij}$. There is a triangle (red dashed lines) $(i,k,j)$, a square (blue dashed lines) $(i,m,l,j)$ and a pentagon $(i,n,o,p,j)$ sharing the edge $e_{ij}$. The black dashed lines represent links to other vertices in the neighborhoods of $i$ and $j$, $N(i)$ and $N(j)$. The independent short-cycle condition is satisfied for the edge $e_{ij}$, but the presence of the square and the triangle depicted would violate the condition for the edges $e_{jk}$ and $e_{ki}$ as they would support a pentagon $(i,m,l,j,k)$ and a triangle $(i,k,j)$ that share two edges. For the edge $e_{ij}$ this graph satisfies the independent short-cycle condition, if and only if either the dashed triangle or square is present, but not both.}
\label{fig:hardcore}
\end{figure}
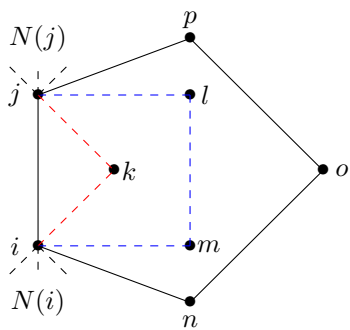

The class of graphs that we shall consider is defined by a network analogue of the statistical mechanical hard core condition.  It is well known that, to avoid an infinite compressibility of a boson gas when lowering the temperature, a hard-core condition must be imposed that keeps the fundamental degrees of freedom, the particles in this case, from overlapping. The independent short-cycle condition \cite{trugenberger2017combinatorial, kelly2019self} is the corresponding network analogue. 
The physical degrees of freedom, which in this case are the triangles, squares and pentagons, can touch (share an edge) but not overlap (share more than one edge). 
Formally, this condition can be defined as,
%
%

\begin{definition}{The independent short-cycle condition.}\label{def:hard_core}
Consider the diagram in Fig. \ref{fig:hardcore}. The independent short-cycle condition is satisfied by a graph $G(V,E)$, if all of its closed cycles of length $n\le 5$ do not share more than one edge (only cycles up to length 5 matter for locality, as explained above). Let $C_n (e_{ij})$ represent a closed cycle supported upon an edge $e_{ij}$ of length $n \le 5$.  
For example in Fig. \ref{fig:hardcore}, $C_4(e_{ij})$ contains the vertices $i,j,l,m$ and is the set $\{ e_{ij},e_{jl},e_{lm},e_{mi} \}$. The independent short-cycle condition is satisfied for the graph $G$ if and only if,
\begin{equation}
    \bigcap\limits_n^{n \le 5} C_n (e_{ij}) = e_{ij} \text{,~} \forall e_{ij} \in E \text{.}
\end{equation}
\end{definition}

As is discussed in the caption, the edge graph depicted satisfies the condition if and only if one of either the dashed square or triangle are present, but not both.
The reader is referred to Kelly {\sl et al} \cite{kelly2019self} where it is shown that it is possible to reduce the condition to a statement excluding certain subgraphs that occur if it is violated.

The physical meaning of the independent short-cycle condition is explained above (for a detailed discussion see \cite{kelly2019self}). Here we
shall focus on the consequences of this condition for the mathematical expressions of graph curvature. In particular, in \cite{kelly2019self} it has been shown that, on graphs that satisfy this condition, the OR curvature reduces to a simple closed form,
\begin{eqnarray}\label{equation:NewOllivCurv}
&& \resizebox{0.44\textwidth}{!}{$\kappa^{\rm OR}(ij)=\frac{\triangle_{ij}} {k_i\land k_j}-\left[1-\frac{1+\triangle_{ij} +\square_{ij}}{k_i\lor k_j}-\frac{1}{ k_i\land k_j}\right]_+ -$}
\nonumber\\
&& \resizebox{0.44\textwidth}{!}{$\left(\frac{\triangle_{ij}}{k_i\land k_j}-\frac{\triangle_{ij}}{k_i\lor k_j}\right)\lor \left(1-\frac{1+\triangle_{ij} +\square_{ij} +\pentagon_{ij}}{k_i\lor k_j}-\frac{1}{k_i\land k_j}\right)\text{,}~$}
\label{ORhardcore}
\end{eqnarray}
where
$k_i$, $k_j$ are the degrees of vertices $v_i$, $v_j$, respectively.
Further the symbols $\triangle_{ij}$, $\square_{ij}$ and $\pentagon_{ij}$ denote the number of triangles, squares and pentagons supported on the edge $e_{ij}$, and  $\alpha\lor\beta:=\max(\alpha,\beta)$, $\alpha \land\beta:=\min(\alpha,\beta)$ and $[\alpha]_+:=0\lor\alpha $ for any $\alpha,\:\beta\in \mathbb{R}$.  Remarkably, the 
independent short-cycle condition allows one to calculate the OR curvature by simply counting the number of short loops based on an edge.

In the following we are interested mainly in the dependence of curvature on {\it global} combinatorial quantities, the numbers of cycles of a given length based on an edge, neglecting local details originating from the different connectivity at the two vertices. 
To expose this dependence we focus on edges which have the same connectivity at their two vertices, by setting $k_i = k_j = \expval{k}$. 
These are the majority of edges when the degree distribution is peaked. 
The local topology of the graph, from the perspective of the edge between $v_i$ and $v_j$, is determined by the cycles it participates in, and, the way in which that subgraph containing the edge is connected into the rest of the graph.
By making this choice we have constrained the variation to be entirely dependent upon cycles, simplifying the problem.
For this case we obtain the even simpler expression,
\begin{eqnarray}
\kappa^{\rm OR}(ij)&&=\frac{\triangle_{ij}} {\expval{k}}-\left[1-\frac{2+\triangle_{ij} +\square_{ij}}{\expval{k}}\right]_+ 
\nonumber \\
&&- \left [1-\frac{2+\triangle_{ij} +\square_{ij} +\pentagon_{ij}}{\expval{k}}\right]_+ \ .
\label{ORmeanfield}
\end{eqnarray}
Of course this expression is exact for all edges of a regular graph.

\section{Forman-Ricci Curvature}
\label{sec:forman}
An alternative discrete measure of graph curvature was introduced by Robin Forman \cite{forman2002combinatorial,forman2003bochner,forman2004topics} using the topological constructs of CW (Closure-finite, Weak) cell complexes.
Forman's work defines an entire parallel apparatus of differential forms, Morse theory and Ricci curvature to those well understood in the traditional algebraic geometry of smooth manifolds.
This richness of structure is intriguing, but unlike the OR curvature described in Section \ref{sec:or_curvature}, there is no direct relationship of the Forman curvature to that of a smooth manifold in which the graph is embedded.
As noted before, in contrast to Regge calculus \cite{regge1961general} and other forms of curvature more commonly used in EDT or CDT, this form of curvature is not defined geometrically.
Its purely combinatorial nature has its origin in the parallel structure of discrete differential geometry elaborated by Forman in his original works.
Given the very different origin of FR curvature to OR curvature, any relationship is not to be expected {\sl a priori}.

The complete treatment of FR curvature is technical and we shall only briefly survey it here.
Essentially it draws upon an analogy with identities developed  by Bochner \cite{bochner1946vector} regarding the decomposition of the Riemannian-Laplace operator on the space of $p$-forms, $\Omega^p(M)$ defined for a manifold $M$.
This decomposition yields a covariant derivative and a curvature correction known as the Bochner-Weitzenb{\"o}ck identity.
Its discrete form is used to derive Forman-Ricci (FR) curvature.

CW complexes (an excellent standard text is Hatcher \cite{hatcher2002algebraic}) are constructed from $p$-cells ($p$  referring to the dimension of the cell).
One constructs a $d$ dimensional CW complex by gluing $p \leq d$ complexes along shared faces.
For our purposes we will focus on cell complexes up to $p=2$, which are essentially equivalent to graphs, with the addition that cycles in the graph are assumed to bound a $2$-cell.
This assumption is critical, and often overlooked in definitions of FR curvature.
In some of the literature, notably \cite{sreejith2016forman}, the inclusion of these $2$-cells is referred to as ``augmented" FR curvature, but we view the non-augmented version as essentially trivial and of no utility in applications to quantum gravity.

We define the boundary of a $p$-cell as the $p-1$ cells that ``contain" the cell.
For a $1$-cell $\langle p_0p_1 \rangle$, the boundary is the collection of points $p_0$ and $p_1$, and for a general $p-$cell, $\alpha_p$, it is a proper face of a $p+1$ cell $\beta$ if it is a member of the boundary set of $\beta$, and we write $\alpha_p < \beta_{p+1}$, or $\beta_{p+1} > \alpha_p$.
A $p$-cell CW complex $M$ over $\mathbb{R}^p$, is defined formally a collection of cells $\alpha_q \ , q \in \{0,\dots,p\}$, such that any two cells are joined along a common proper face, and all faces are contained in the cell complex.

An important concept when developing the  curvature of cell complexes is the definition of the neighbors of a given $p$-cell  \cite{forman2002combinatorial,forman2003bochner} introduced by Forman as,

\begin{definition}\label{def:neighbor}
	$\alpha_1$ and $\alpha_2$ are $p$-cells of a complex $M$. $\alpha_1$,$\alpha_2$ are neighbors if:
	\begin{enumerate}
		\item $\alpha_1$ and $\alpha_2$ share a $(p+1)$ cell $\beta$ such that $\beta > \alpha_1$ and $\beta > \alpha_2$, or
		\item $\alpha_1$ and $\alpha_2$ share a $(p-1)$ cell $\gamma$ such that $\gamma < \alpha_1$ and $\gamma < \alpha_2$.
	\end{enumerate}
\end{definition}

Further, we can partition the set of neighbors of a cell into parallel and non-parallel.
Two $p$-cells $\alpha_1$,$\alpha_2$ are parallel neighbors, if one but not both of the conditions in Def. \ref{def:neighbor} are true, and write $\alpha_1 \parallel \alpha_2$.

With these concepts, FR curvature is defined as a series of maps $\mathcal{F}_p : \alpha_p \rightarrow \mathbb{R}$, for each value of $p$, and has the following simple form,

\begin{equation}\label{eqn:fr-simple}
	\mathcal{F}_p (\alpha_p) = \# \{ \beta_{(p+1)} > \alpha_p \} + \# \{ \gamma_{(p-1)} < \alpha_p \} - \# \{ \epsilon_q \parallel \alpha_p \} \mbox{,} 
\end{equation}
where $\epsilon_q$ is a $q$-cell that is a parallel neighbor of $\alpha_p$, and $q \neq p$.
The symbol $\#$ is intended to denote the number of such cells satisfying the condition in braces.
Essentially this definition computes $\mathcal{F}_p (\alpha_p)$ as the number of $p-1$-cells that bound $\alpha_p$, plus the number of $p+1$ cells of which $\alpha_p$ is part of the boundary minus the number of parallel neighbors of $\alpha_p$.

It is possible to adorn each cell with a weight, and for completeness we reproduce here the full version of this formula for weighted complexes.
The addition of weights to the cells is of particular interest when introducing discrete analogs of Morse theory and differential forms \cite{forman1995discrete,forman2004topics,forman2002combinatorial}, where the weights are used to identify critical points in the topology and define gradient vector fields.
For each $p$-cell $\alpha_p$ we associate a weight $g_{\alpha}$ and we denote by $\tilde{\alpha_p}$ its neighbors per Def. \ref{def:neighbor}.
Using these definitions we have,
\begin{multline}\label{eqn:fr-full}
	\mathcal{F}_p(\alpha)=g_{\alpha} \left \{ \sum\limits_{\beta_{(p+1)} > \alpha } \frac{g_{\alpha}}{g_{\beta}}	+  \sum\limits_{\gamma_{(p-1)} < \alpha } \frac{g_{\gamma}}{g_{\alpha}}\right. \\
	\left. - \sum\limits_{\tilde{\alpha_p} \parallel \alpha_p }\abs{\left[ \sum\limits_{ \substack{ \beta_{(p+1)} > \alpha_p\\ \beta_{(p+1)}> \tilde{\alpha}_p }} \frac{\sqrt{ g_{\alpha_p} g_{\tilde{\alpha}_p} } }{ g_{\beta}}  - \sum\limits_{ \substack{ \gamma_{(p-1)} < \alpha_p\\ \gamma_{(p-1)} < \tilde{\alpha}_p }} \frac{ g_{\gamma}}{\sqrt{ g_{\alpha_p} g_{\tilde{\alpha}_p} } }  \right] } \right \}
\end{multline}
It is important to note that the last term in this equation is a sum over the absolute difference between $p+1$ and $p-1$ parallel neighbors.
For an unweighted graph this reduces to Eq. \eqref{eqn:fr-simple}.

If we set $p=1$, Forman identifies Eq. \eqref{eqn:fr-simple} as the Ricci curvature, defined on the edges of an unweighted graph, and we refer to this as the Forman-Ricci (FR) curvature. We distinguish this particular value  of the FR curvature for $p=1$ by the notation $\kappa^{F}_{ij} = \mathcal{F}_1(e_{ij})$.
For a graph Eq. \eqref{eqn:fr-simple} is simple; the vertices and edges constitute the $0$ and $1$ cells, and closed loops in the graph constitute the $2$-cells.
It is this simplicity that underlies the favorable computability of FR curvature.

The FR curvature is a combinatorial quantity and, as is evident from its definition in Eq. \eqref{eqn:fr-simple}, there is no arbitrary restriction to the length of cycles that are admissible as bounding a $2$-cell. The so called ``augmented" Forman curvature is obtained by restricting these cycles to triangles, although this restriction is not present in the original work of Forman \cite{forman2003bochner}. However, as we explained above, this is not appropriate for applications to CQG, where discrete locality requires taking into account cycles of length up to $5$. 
We define thus an appropriate ``enhanced" Forman curvature $\kappa^{FR}$ by truncating the expansion to cycles of length $5$, and note that Def. \ref{def:hard_core} similarly covers cycles up to length $5$. 

Assuming that a graph possesses the independent short-cycle property \cite{kelly2019self} brings a substantial simplification also for the enhanced Forman curvature. Indeed, in this case we can write the following simple expression for $\kappa^{FR}_{ij}$ that is exact if the graph does not have closed cycles larger than pentagons,

\begin{equation}\label{eqn:exact_fr_expansion}
    \kappa^{FR}_{ij}=4-k_i -k_j +3 \triangle_{ij} + 2 \square_{ij} + \pentagon_{ij} \text{,}
\end{equation}
where, for a given edge either $\triangle_{ij} > 0$ or $\square_{ij}>0$, but not both. The importance of the independent short-cycle condition is that it allows one to compute the contribution to the number of parallel edges to $e_{ij}$ from the edges ($1$-cells) that are not part of any cycle incident upon $e_{ij}$.
This is easily obtained, as the independent short-cycle condition implies that every cycle incident upon a given edge consumes precisely one edge from each of the vertices $v_i$ and $v_j$.
Inspecting Eq. \eqref{eqn:fr-simple} the role that this condition plays can be understood by taking each term in turn.
The first term simply counts the number of short cycles on the edge ($\triangle_{ij}+\square_{ij} +\pentagon_{ij}$) with the second term always contributing $2$, being the number of $0$-cells or vertices per edge.
The third term requires more care, and we note that when the independent short-cycle condition is satisfied we can again use the number of short cycles to compute the number of parallel edges. 
Specifically each triangle contributes $0$ parallel edges, each square $1$ and each pentagon $2$.
We can then subtract from the degrees of each vertex of the edge those edges participating in a short cycle, and of course the edge connecting the vertices.
As such, the number of parallel edges that share a vertex with a given edge $e_{ij}$ is precisely $k_i + k_j - 2 - 2(\triangle_{ij} + \square_{ij} + \pentagon_{ij})$. To arrive at Eq. \eqref{eqn:exact_fr_expansion}, we add this to the first term and obtain our final result.

Eq. \eqref{eqn:exact_fr_expansion} further simplifies for edges with the same connectivity $\expval{k}$ at their two vertices, as we have considered for the OR curvature in \eqref{ORmeanfield},
\begin{equation}\label{eqn:fr_mean_field}
    \kappa^{FR}_{ij}= 4-2\expval{k} + 3 \triangle_{ij} +2 \square_{ij} +\pentagon_{ij}  \text{.}
\end{equation}
In this case, this expression can be considered as a mean field approximation for the FR curvature in a consistent manner to that used in the OR curvature.
We have essentially replaced the local degrees of the vertices with an average to avoid the details of their local connectivity, whilst retaining the global combinatorial measures as the number of short cycles incident upon the edge.
When this expression is summed over all edges to obtain the equivalent of the Ricci scalar for the graph it becomes exact. As we will see, for OR curvature this is not so since the proper treatment of the $[]_+$ terms in Eq. \eqref{ORmeanfield} requires correction terms to be applied to the averaged curvature sum.

\section{Comparing Ollivier-Ricci and Forman-Ricci Curvature}
Considering Eq. \eqref{ORmeanfield}, the effect of the $[]_+$ in the last two terms makes a direct comparison of this expression with the FR curvature difficult.
To make this comparison easier we express the OR curvature in Eq. \eqref{ORmeanfield} as a ``mean field" term, defined as the expression in Eq. \eqref{ORmeanfield} in which all brackets are simply summed up without taking into account the ``+" subscripts, plus a correction term. The mean field term represents the sought after dependence on the numbers of cycles alone, while the correction embodies deviations dependent on connectivity (and, as explained above there can be further corrections dependent on other connectivity differences).
This gives 
\begin{eqnarray}
    \expval{k} \kappa^{OR}_{ij} &&= \expval{k} \kappa^{ORMF}_{ij}+ \delta_{ij} \ ,
    \nonumber \\
    \kappa^{ORMF}_{ij} &&= 4-2\expval{k} +3 \triangle_{ij} + 2\square_{ij} +\pentagon_{ij} \ ,
\end{eqnarray}
where the correction to the mean field value $\kappa^{ORMF}_{ij}$ is given by Eq. \eqref{eqn:mf_cases}.

\begin{widetext}
\begin{equation}\label{eqn:mf_cases}
\delta_{ij} = \begin{cases}
0 &\text{if $\expval{k} > 2+ \triangle_{ij}+\square_{ij}+\pentagon_{ij}$}\\
\expval{k}-2-\triangle_{ij}-\square_{ij}-\pentagon_{ij} &\text{if $2+\triangle_{ij}+\square_{ij} < \expval{k} < 2+ \triangle_{ij}+\square_{ij} + \pentagon_{ij}$} \\
2\expval{k}-4-2\triangle_{ij}-2\square_{ij}-\pentagon_{ij} &\text{if $ \expval{k} < 2+\triangle_{ij}+\square_{ij}$}
\end{cases}
\end{equation}
\end{widetext}

This shows that the properly enhanced local Forman Ricci curvature coincides (up to an overall factor) with the mean field value of the Ollivier Ricci curvature,
\begin{equation}\label{eqn:mean_field_equivalence}
\expval{k} \kappa^{ORMF}_{ij} = \kappa^{FR}_{ij} \ .
\end{equation} 
This result is surprising, given that the two discrete curvature constructions have completely different origins. It is a first indication that at least the global combinatorial dependence on the number of cycles is unique for any local discrete curvature measure. 

The correction term vanishes and the two curvatures become essentially identical for graphs with large connectivity and sparse cycles. Unfortunately this is not the relevant case for applications to CQG, where the emergence of geometry is associated with a condensation of short cycles. 
Indeed, it has been shown in a series of papers that the emergence of geometry from random graphs is related to clustering, the appearance of large numbers of triangles, in the curved case \cite{krioukov2016clustering} and with the appearance of large numbers of squares, 4-cycles, in the flat case  \cite{trugenberger2017combinatorial, kelly2019self}. In this flat case, the geometric ground state is a torus lattice, a periodic hyper-cubic mesh. In this case both measures coincide and are numerically zero, matching with the intuitive interpretation of this state as a Ricci flat ground state.
For this very restricted case the inequality and Eq. \eqref{eqn:mean_field_equivalence} is trivially satisfied, but even the presence of a single additional edge bisecting a square would violate the conditions under which the equality is valid.

To further test this rather surprising result, we can generate a variety of random graphs, with varying connectivity and edge density, and compute both curvature values to compare the results.
For the purposes of this simulation we have used Erd{\"o}s-R{\'e}nyi (ER) random graphs \cite{barabasi2016network} (in fact, as we are using a fixed graph size they should be correctly termed Gilbert graphs), using a varying link probability $p$ in the range $0.01$ to $0.1$, after which we manually enforce the independent short-cycle condition by removing edges that violate it.
The mechanism for the removal of edges that violate the independent short-cycle condition is conducted by checking each edge in the graph for its participation in a well defined short list of subgraphs that indicate violation.
This list of subgraphs is taken from Fig. 1 of \cite{kelly2019self}.
A similar analysis could be conducted with other random graph types such as Watts-Strogatz \cite{watts1998collective} or Barab{\'a}si-Albert \cite{albert2002statistical}.
The choice of ER graphs is specifically taken as they can easily be constructed to arbitrary edge density by increasing the value of $p$.
Our aim is to understand how the curvature measures behave as the conditions in Eq. \eqref{eqn:mf_cases} are progressively violated, and as such ER graphs are sufficient.

\begin{figure}[ht]
	\centering
	\includegraphics[scale=0.43]{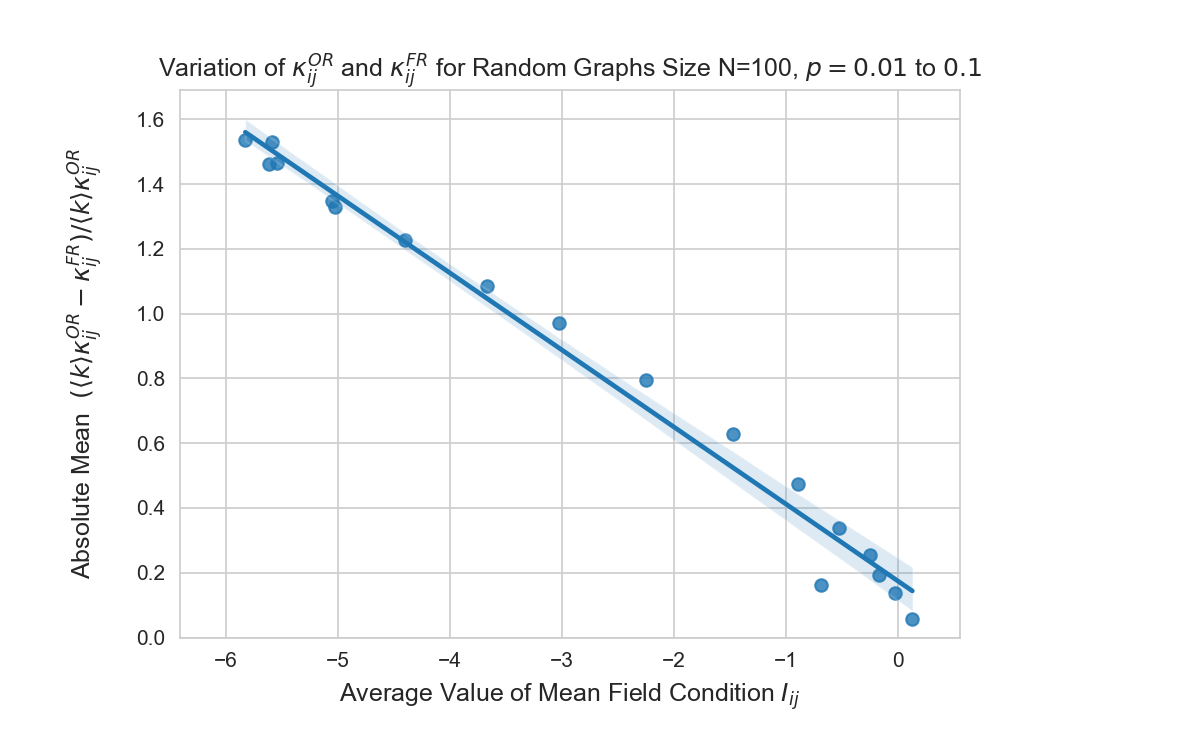}
	\caption{For $N=100$ vertices we generate random Erd{\"o}s-R{\'e}nyi graphs, varying the link probability $p$. For a collection of $10$ graphs for each value of $p$, we compute the average absolute fractional difference between $\expval{k} \kappa^{OR}_{ij}$ and $\kappa^{FR}_{ij}$, which we plot against the mean-field condition $I_{ij}=\expval{k}-2-\triangle_{ij}-\square_{ij} -\pentagon_{ij}$ from Eq. \eqref{eqn:mf_cases}. The blue dots represent the raw simulation values, with the blue line representing the linear regression fit, shading to cover the 95\% confidence interval for the regression.}
	\label{fig:ScalarComparison}
\end{figure}

For a fixed number $N=100$ of vertices this generates graphs with an average degree $\expval{k}>1$ above the critical threshold for the emergence of a large connected subgraph.
Although the generation of random graphs is computationally easy, computing the curvature values (particularly OR curvature) is not.
It is for that reason that we have not conducted our analysis for much bigger graphs, but we did not find significant differences between graphs up to $N=100$.
As $p$ increases, the edge density of the graph will also increase, along with the density of short cycles.
For each edge we compute $I_{ij}=\expval{k}-2-\triangle_{ij}-\square_{ij} -\pentagon_{ij}$. From Eq. \eqref{eqn:mf_cases}, when $I_{ij}>0$ the two measures of curvature should be identical up to the factor of $\expval{k}$.
In Fig. \ref{fig:ScalarComparison} we plot the average fractional difference between the two curvature measures, $(\expval{k} \kappa^{OR}_{ij} - \kappa^{FR}_{ij})/\expval{k} \kappa^{OR}_{ij}$ over all edges in randomly generated graphs against the average value of $I_{ij}$.
For each link probability $p$ we generate $10$ graphs (the choice of $10$ being a computational complexity limit), to avoid the results being skewed by unusual graph configurations, and we compute the OR curvature using the  NetworkX python toolkit, as extended by Chien-Chun Ni \cite{hagberg2008exploring,ni2019community}.
For  FR curvature we use our own library, as the publicly available libraries do not include longer cycles.
We plan to make available this extension publicly, but the code is available on request from the authors.
 
The simulation clearly shows that the two curvature measures differ when $I_{ij}<0$ but, as $I_{ij}$ increases and approaches zero they converge to the same value. 
This is fully consistent with our analysis and provides supporting numerical evidence for our main claim.

%
%

\section{Conclusion}
\label{sec:conclusion}

EDTs are appealing because singularities typically associated with Lorentz metrics can be avoided, but are hard to reconcile with the correct semi-classical ground state. CDTs have much better properties in this respect, but, in our opinion, they have the disadvantage to assume a Lorentz structure on all scales {\sl ab initio}. The CQG program is aimed to construct a bridge between the two. The idea, or better hope at this moment, is that the observed Lorentzian universe emerges only at scales larger than a critical scale below which all physics is Euclidean. 
The aim of this work was to explore how two very different measures of discrete curvature of graphs used in this program are related.
This question has important ramifications for models of CQG and emergent geometry.
In particular, the OR curvature has a rich correspondence to traditional Riemann-Ricci measures of curvature in smooth manifolds, and any theory of discrete quantum gravity needs to be able to connect with such concepts in a low energy continuum limit.
On the other hand the FR curvature, defined in a strictly combinatorial setting, has no such connection, but has the advantage of being somewhat easier to compute and is associated with an entire parallel structure of index theories and discrete differential geometry.
Our proof that these two measures are related to each other, even if for a restricted set of graphs, is a key step forward in the CQG program.

\bibliographystyle{eplbib}
\bibliography{CurvatureEquivalence}

\end{document}